\documentclass[%
 reprint,
 amsmath,amssymb,
 aps,
]{revtex4-1}

\usepackage{graphicx}
\usepackage{dcolumn}
\usepackage{bm}

\newcommand{\be}{\begin{equation}}
\newcommand{\ee}{\end{equation}}
\newcommand{\ba}{\begin{eqnarray}}
\newcommand{\ea}{\end{eqnarray}}
\newcommand{\baa}{\begin{eqnarray*}}
\newcommand{\eaa}{\end{eqnarray*}}
\newcommand{\bb}{\begin{thebibliography}}
\newcommand{\eb}{\end{thebibliography}}

\newcommand{\bi}[1]{\bibitem{#1}}
\newcommand{\lab}[1]{\label{#1}}
\newcommand{\re}[1]{(\ref{#1})}

\renewcommand\t{\tilde}

\newcommand\ve{\varepsilon}

\newtheorem{ver}{Version}

\newcounter{my}
\newcommand{\he}%
   {\stepcounter{equation}\setcounter{my}%
   {\value{equation}}\setcounter{equation}0%
   }%
\newcommand{\she}%
   {\setcounter{equation}{\value{my}}%
    }%

\begin{document}

\preprint{APS/123-QED}

\title{Almost perfect state transfer in quantum spin chains}

\author{Luc Vinet}
 \affiliation{Centre de recherches math\'ematiques
Universit\'e de Montr\'eal, P.O. Box 6128, Centre-ville Station,
Montr\'eal (Qu\'ebec), H3C 3J7}
\author{Alexei Zhedanov}%
\affiliation{Institute for Physics and Technology, R.Luxemburg
str. 72,  83114, Donetsk, Ukraine }


\begin{abstract}
The natural notion of almost perfect state transfer (APST) is examined. It is applied to the modelling of efficient quantum wires with the help of $XX$ spin chains. It is shown that APST occurs in mirror-symmetric systems, when the 1-excitation energies of the chains are linearly independent over rational numbers. This result is obtained as a corollary of the Kronecker theorem in Diophantine approximation.  APST happens under much less restrictive conditions than perfect state transfer (PST) and moreover accommodates the unavoidable imperfections. Some examples are discussed.

\end{abstract}

\pacs{}

\keywords{perfect state transfer, spin chains, orthogonal
polynomials}

\maketitle

\section{Introduction}
\setcounter{equation}{0}
The design of models to transfer quantum states between distant locations is of relevance to many quantum information protocols. In recent years, spin chains have been proposed \cite{Bose}, \cite{Albanese} as possible basic systems in the construction of such quantum channels. In these devices, the state of the qubit at one end of the chain is transfered to the qubit at the other end after some time. A key advantage of this method is that it minimizes the need for external interventions as the transfer is realized through the intrinsic dynamics of the chain.

The efficiency and reliability of quantum wires must be high. Ideally, it is wished that the probability of finding the initial state as the  output be one; when this is so, one speaks of perfect state transfer (PST). It has been shown, in particular, that PST can be achieved in spin chains by properly engineering and modulating the couplings between the sites \cite{Albanese}, \cite{Christ}.

In this context, a question of  practical importance is that of the robustness of these ideal transfer properties in view of the unavoidable manufacturing  and experimental deviations from the theoretical specifications.

There are many sources of errors: nonsynchronous or imperfect input and read-out operations, fabrication defects, additional interactions, systematic biases etc. What their influences on the fidelity of state transfer are \cite{Kay1}, \cite{RSA} and how to correct or circumvent them \cite{BB} , \cite{MKE}  has been the object of various studies.     We shall here relate particularly to the errors imputable to the quantum wires and measurements.  In \cite{fractal} (see also \cite{bounds}, \cite{disorder}) the imperfections in the production of the device are modeled by adding random perturbations to the couplings and magnetic fields of a chain with PST.   In \cite{Bruderer} (which uses methods similar to \cite{Kay} and \cite{VZ_PST}), modulated chains are coupled to boundary states to make transfer more robust against imperfections which are also randomly simulated. We shall adopt in the following a complementary approach.

We take for given that the manufacturing of the chains will not be perfect and that the precise PST requirements will not be met. Notionally, we assume that the imperfections are static. We then consider under which conditions will state transfer be almost perfect. In other words, instead of examining the effect of perturbations on PST chains, we shall readily attempt to characterize the chains that somehow incorporate defects and come "very" close to achieving PST.

The simplest spin chains exhibiting PST are governed by XX Hamiltonians with nearest-neighbor interactions. We shall confine ourselves  to these systems in the following. Given that the total spin projection commute with the Hamiltonians, many of the state transfer properties can be obtained by focusing on the 1-excitation sector of the state space. The corresponding restrictions of the Hamiltonians are tri-diagonal Jacobi matrices $J$ whose entries are the couplings and magnetic fields of the chains. Naturally, these $J$ are diagonalized by orthogonal polynomials (OPs). The perfect transfer of a single spin up from one end of the chain to the other is possible only if $J$ is mirror symmetric; moreover it puts strong requirements on its spectrum.   The theory of orthogonal polynomials is very useful to obtain these results.

The same framework will be adopted to determine the conditions for almost-perfect state transfer APST. The question now is: under what circumstances are there times for which the transition probabilities from one site to another can be made to approach 1. Mirror symmetry will again be necessary as we shall see. It will be sufficient however for the spectrum of a mirror-symmetric $J$ to obey conditions much milder than in the PST case.  Interestingly, these results will follow from the Kronecker theorem in Diophantine approximation.

The remainder of the paper will proceed as follows. We shall first review in Section 2  how 1-excitation state transfer is realized in $XX$ spin chains and how the eigenstates are obtained with the help of OPs. Almost perfect state transfer will be defined in Section 3 and necessary and sufficient conditions for its occurrence will be determined. In Section 4, it will be shown how various chains with APST can be obtained from a parent chain with APST through a procedure referred to as spectral surgery \cite{VZ_PST}. Section 5 offers a number of illustrative examples: situations where APST occurs and cases where neither PST nor APST are possible. In a significant model, the times for which APST is attained can be estimated showing in this instance that a good approximation to PST is obtained in finite time independently of the size of the chain.

Let us finally draw attention to recent related and complementary publications that have made use of the same tools from number theory in the case of spin chains with uniform couplings. In his thesis  \cite{Burgarth}, using the Kronecker theorem,  Burgarth has shown that the Heisenberg XX spin chains with prime lengths have arbitrary good fidelity. In \cite{Godsil}, Godsil introduced the notion of pretty good state transfer on graphs which is equivalent to our definition of APST. Our terminology is dictated by the fact that in these instances the transition amplitudes are almost periodic functions. Recently, the characterization of the uniform $XX$ Heisenberg chains which admit this pretty good state transfer was analyzed using again number theoretic methods \cite{Godsil1}.

In dealing with the Heisenberg $XX$ spin chain with uniform coupling and zero magnetic fields, the mirror-symmetry of the one-excitation Hamiltonian is de facto ensured as this operator coincides with the adjacency matrix of the $(N+1)$-path. As will be seen, this symmetry requirement comes into play in relation with APST, when more general couplings and magnetic fields are considered. It is remarked in \cite{Godsil} that in order to get a good approximation to perfect state transfer, the waiting time must be very large. Numerical results presented in \cite{Godsil1}, indicate that for a given level of fidelity, or approximation, the required times grow linearly with $N$, the numner of sites in the chain minus one. (We shall comment on these observations in Section 5, using exact results on almost perfect return \cite{GVZ} as applied to the uniform $XX$ chain.)   Thus Godsil \cite{Godsil} expressed the view that pretty good state transfer or APST will not be a satisfactory substitute for perfect state transfer in practice. We here wish to stress in this connection that when considering non-uniform couplings, there are situations, as shown by an aforementioned example in Section 5, where high fidelity can be achieved in finite time irrespectively of the chain length. This suggests that the class of spin chains with APST, a much larger one than the PST family, should not be so radically deemed of impractical use at this point.

\section{1-Excitation state transfer in $XX$ spin chains}
\setcounter{equation}{0}

We shall consider $XX$ spin chains
with nearest-neighbor interactions.  Their Hamiltonians $H$ are of the form
 \be H=\frac{1}{2} \:
\sum_{l=0}^{N-1} J_{l+1}(\sigma_l^x \sigma_{l+1}^x + \sigma_l^y
\sigma_{l+1}^y) +  \frac{1}{2} \: \sum_{l=0}^N B_l(\sigma_l^z +1),
\lab{H_def} \ee where $J_l$ are the constants coupling the sites
$l-1$ and $l$ and $B_l$ are the strengths of  the magnetic field
at the sites $l$ ($l=0,1,\dots,N$). The symbols $\sigma_l^x, \:
\sigma_l^y,\: \sigma_l^z$ stand for the Pauli matrices which act
on the $l$-th spin.

It is immediate to see that
$$
[H, \frac{1}{2} \: \sum_{l=0}^N (\sigma_l^z +1)]=0
$$
which implies that the eigenstates of $H$ split in subspaces
labeled by  the number of spins over the chain that are up.
In order to characterize the chains with APST, it will suffice to restrict $H$
to the subspace spanned by the states witch contain only one
excitation. A natural basis for that subspace is
given by the vectors
$$
|e_n \rangle = (0,0,\dots, 1, \dots, 0), \quad n=0,1,2,\dots,N,
$$
where the only "1" occupies the $n$-th position.  The restriction $J$ of $H$ to the 1-excitation subspace acts as follows
\be J |e_n \rangle  = J_{n+1} |e_{n+1} \rangle + B_n |e_n \rangle + J_{n}
|e_{n-1} \rangle. \lab{Je} \ee Note that \be
J_0=J_{N+1}=0 \lab{J0} \ee is assumed.

Consider the polynomials $\chi_n(x)$ obeying the recurrence relation
\be J_{n+1} \chi_{n+1}(x) + B_n \chi_n(x) +
J_{n} \chi_{n-1}(x) = x \chi_n(x) \lab{rec_chi} \ee
with
\be
\chi_{-1}=0, \; \chi_0=1 . \lab{ini_chi} \ee
They satisfy the orthogonality relations
\be
\sum_{s=0}^N w_s \chi_n(x_s) \chi_m(x_s) = \delta_{nm}, \lab{ort_pi} \ee
where $w_s$ are the discrete weights taken to satisfy
\be
\sum_{s=0}^N w_s =1. \lab{norm_w} \ee

In what follows we shall take the eigenvalues $x_s$ in increasing order
\be
x_0<x_1<x_2 < \dots <x_N . \lab{incr_x} \ee
Let
\be
| x_s \rangle = \sum_{n=0}^N \sqrt{w_s} \chi_n(x_s) |e_n \rangle \ . \lab{sew_expans} \ee
It is easily seen that these vectors $|s \rangle$ are eigenstates of $J$ with eigenvalues $x_s$
\be
J |x_s \rangle = x_s |x_s\rangle . \lab{J_s} \ee
Since both bases $\{|s\rangle\}$ and $\{| e_n\rangle\}$ are orthonormal
$$
\langle e_n|e_m \rangle =\delta_{nm}, \quad  \langle x_{s'} | x_s\rangle =\delta_{ss'},
$$
we also have
\be
| e_n \rangle = \sum_{s=0}^N \sqrt{w_s} \chi_n(x_s) |x_s \rangle . \lab{esw_expans} \ee

Let $P_{N+1}(x)$ be the characteristic polynomial of $J$  \be
P_{N+1}(x) = (x-x_0)(x-x_1) \dots (x-x_N).  \lab{P_N+1} \ee The discrete weights are expressed as \cite{Chi}
\be w_s =
\frac{h_N}{P_N(x_s) P_{N+1}'(x_s)}, \quad s=0,1,\dots, N
\lab{w_s_PP} \ee
with $h_N=J_1^2 J_2^2 \dots J_N^2$ and $P_N(x) =h_N^{1/2}\chi_N(x)$.

\section{Necessary and sufficient conditions for almost perfect quantum state transfer}
By "almost perfect state transfer" (APST) we mean the following.

Assume that the spin chain is prepared at time $t=0$ in the pure state $|e_0 \rangle$. This means that that at site $n=0$ the spin is up while all other spins are down.

For arbitrary times $t$, this state will evolve into the state
\be
| e_0(t)\rangle  = e^{-i t J} |e_0 \rangle .
\lab{pqc} \ee
We demand that for any small $\epsilon>0$, there exists a value of time $t$ such that
\be
\left| | e_0(t) \rangle - e^{i \phi(t)} | e_N \rangle  \right|^2 < \epsilon, \lab{main_lim} \ee
where $\phi(t)$ is a real parameter which can depend on $t$. This means that the state $| e_0(t) \rangle$ can be as close to the state $| e_N \rangle$ as desired and that $| e_0 \rangle$ has thus undergone at time $t$ an almost perfect transfer.

The notation $|\eta - \xi|^2$ for two vectors $\xi$ and $\eta$ stands as usual for
\be
|\xi - \eta|^2 = \sum_{k=0}^N |\xi_k - \eta_k|^2  \lab{xi_eta} \ee
where $\xi_k$ are the expansion coefficients of the vector $\xi$ over a basis, say  $| e_k \rangle$:
\be
\xi = \sum_{k=0}^N \xi_k | e_k \rangle . \lab{xi_k} \ee

Recall that the condition for 1-excitation PST reads
\be
| e_0(t)\rangle = e^{-i \phi(t)} | e_N \rangle ; \lab{PST} \ee
this means that there then exists a time $t$ for which  the state   $| e_0(t)\rangle$ coincides (up to a phase factor $e^{i \phi(t)}$) with the state $| e_N \rangle$.

In the case of almost perfect state transfer, there is no time $t$ for which condition \re{PST} is verified. Nevertheless, it is possible to approach  the state $|e_N \rangle$ with any prescribed degree of accuracy. From a practical point of view, there is no essential difference between perfect and almost perfect state transfer, owing to inevitable technological and measurement  errors. However, the APST conditions are much weaker than the PST ones. APST therefore widens the possibilities for constructing efficient quantum wires.

Let us first derive the necessary condition for APST.

Taking into account expansion \re{esw_expans}, we have
\be
| e_0(t)\rangle = \sum_{s=0}^N \sqrt{w_s} e^{-i x_s t} | x_s \rangle, \quad   | e_N \rangle = \sum_{s=0}^N \sqrt{w_s} \chi_N(x_s) | x_s \rangle \lab{0N_expan} \ee
and hence
\ba
&&\left| e^{-i \phi(t) } |e_0(t)\rangle - |e_N \rangle \right|^2 = \nonumber \\
&&\sum_{s=0}^N  w_s \left|e^{-i \phi(t) - i t x_s} - \chi_N(x_s)\right|^2 . \lab{diff_0N} \ea
It is easily seen that in order to fulfill condition \re{main_lim} we need
\be
|\chi_N(x_s)| =1 . \lab{con_chi_1} \ee
To convince oneself of this fact, suppose that  \re{con_chi_1} does not hold and assume that for some $s=0,1,\dots, N$, we have $\chi_N(x_s)=a$ with $|a| \ne 1$. In this case, using the reverse triangle inequality we have 
$$
\left|e^{-i \phi(t) - i t x_s} - \chi_N(x_s)\right|^2 \ge (|a|-1)^2 >0
$$
and as a consequence the r.h.s. of \re{main_lim} cannot be made arbitrarily small.

Since $\chi_N(x)$ has only real coefficients \re{con_chi_1} implies that $\chi_N(x_s) = \pm 1$. From general properties of orthogonal polynomials \cite{Chi}, \cite{VZ_PST} it follows, using \re{P_N+1}, that  \re{con_chi_1} amounts to
\be
\chi_N(x_s) = (-1)^{N+s}. \lab{chi_N_s} \ee
As shown in \cite{VZ_PST}, \re{chi_N_s} implies  that the matrix $J$ is persymmetric, or mirror-symmetric i.e.
\be
RJR =J, \lab{persym} \ee
where $R$ is the reflection matrix
\[
R=\begin{pmatrix}
  0 & 0 & \dots & 0 & 1    \\
  0 & 0 & \dots  & 1 & 0  \\
    \dots  & \dots & \dots & \dots & \dots      \\
   1 & 0 &  \dots & 0 &0  \\

\end{pmatrix}.
\]
We have thus obtained a necessary condition for APST namely, that the Jacobi matrix $J$ corresponding to the spin $XX$ Hamiltonian should be mirror symmetric. This coincides with one of the necessary conditions for PST \cite{Kay}. The other condition for PST requires that
\be
x_{s+1}-x_s=\frac{\pi}{t} M_s, \lab{PST_2} \ee where $M_s$ are arbitrary positive odd numbers.

We shall now obtain the conditions on the spectrum of $J$  for APST.

The matrix $J$ is Hermitian and mirror symmetric with $J_i>0$ and hence all its eigenvalues $x_s$ are real and distinct.

Let $f_{0n}(t)$  be the amplitude for finding the system in the state $| n \rangle $ at time $t$ if was  in the state $| e_0 \rangle$ at time $t=0$,
\be
f_{0n}(t) = \langle e_n | e^{-iJt} | e_0 \rangle . \lab{amp_f} \ee
It is easily seen that \cite{VZ_PST}
\be
f_{0N}(t) = \sum_{s=0}^N w_s e^{-ix_s t} \chi_N(x_s) . \ee
Taking into account \re{chi_N_s} we have
\be
f_{0N}(t) = \sum_{s=0}^N w_s e^{-ix_s t} (-1)^{N+s}. \lab{f_0N} \ee

Mindful of the normalization \re{norm_w}, it is seen that condition $|f_{0N}(t)| \approx 1$ is equivalent to the condition that 
\be
e^{-ix_s t} (-1)^{N+s} \approx e^{i \phi} \lab{approx_phi} \ee
for a fixed value $t$, where $\approx$ means "approximately equal with any prescribed accuracy". To paraphrase \re{approx_phi}, there thus should be a value $t$ of time for which the l.h.s. is as close as desired to a phase independent of $s$.

This means that $|f_{0N}(t)|$ is an almost-periodic function.

Recall that any almost periodic function $f(t)$ is a formal trigonometric series \cite{besic}
\be
f(t) = \sum_{n=-\infty}^{\infty} a_n e^{i \omega_n t},
\lab{apfdef} \ee
where $\omega_n$ are real parameters. For periodic functions $f(t+T)=f(t)$, one has $\omega_n = \frac{2 \pi n}{T}$ and         \re{apfdef} is the ordinary Fourier series. For almost periodic functions there exist the so-called almost periods. This means that for every $\ve>0$ there exists a real parameter $T=T(\ve)$ such that the inequality
\be
|f(t+T)-f(t)| < \ve
\lab{almost_T} \ee
holds for all $t$.

In turn, condition \re{approx_phi} is tantamount to an inequality involving the exponents that can be stated as follows:

For every $\delta>0$ there exist a real parameter $\phi$ and a value of time $t$ such that 
\be
|x_s t - \pi s + \phi| < \delta \quad \mbox{(mod} \;  2 \pi). \lab{cond_x_s} \ee
In more details, \re{cond_x_s} can be rewritten in the form
\be
- \delta < x_s t - \pi s + \phi + 2 \pi M_s < \delta, \lab{cond_x_s_2} \ee
where $M_s$ are integers which may depend on $s$.

This obviously amounts to a condition on the spectrum of $J$ for APST to occur. The question is: what are the properties that the eigenvalues $x_s$ must possess to ensure that it is possible to find a time $t$ and integers $M_s$ so that \re{cond_x_s_2} is satisfied and hence APST realized. In other words, given the set of real numbers $a_s=\phi-\pi s$, this is asking: when is it possible to find values of $t$ for which $x_s t$ is approximated in terms of integers by $a_s - 2 \pi M_s$ with any prescribed accuracy?

The solution of this Diophantine approximation problem \re{cond_x_s_2} is given remarkably  by the Kronecker theorem \cite{HW}, \cite{LZ}.

In order to state it, let us introduce the following definition \cite{HW}, \cite{LZ}: A set $\alpha_i, i=1,2,\dots $ of real numbers is called linearly independent if for any $n$ the only rational values of $r_1, \dots, r_n$ satisfying
$$
r_1 \alpha_1 + r_2 \alpha_2 + \dots + r_n \alpha_n=0
$$
are $r_1=r_2= \dots = r_n=0$.

The Kronecker theorem can be formulated in two versions, one a special case of the other. The first version is attributed to Kronecker himself.   

\begin{ver}
Assume that the real numbers $x_s, \: s=0,1,\dots N$ are linearly independent over the field of rational numbers. Let $a_0, a_1, \dots, a_N$ be fixed arbitrary real numbers. The Kronecker theorem states that for every $\delta>0$ there exist a real $t$ and integers $M_0, M_1, \dots M_N$ such that the inequalities
\be
|x_s t - a_s - 2 \pi M_s| < \delta, \quad s=0,1,2,\dots, N \lab{Kron} \ee
hold.
\end{ver}

We see how this theorem directly applies to the APST problem. Indeed, provided that the eigenvalues are linearly independent  over the rational numbers, we may conclude from the Kronecker theorem that for every parameter $\phi$ it is possible to find a time $t$ so that $|f_{0N}(t)|$ is as close to 1 as desired.

In many cases however, the eigenvalues $x_s$ are not linearly independent. This means that there can exist $L$ independent relations of the type
\be
r_0^{(i)} x_0  + r_1^{(i)} x_1 + \dots + r_N^{(i)} x_N=0, \quad i=1,2,\dots, L \le N
\lab{int_rel} \ee
where $r_s^{(i)}$ are integers such that for every $i=1,2,\dots, L$ at least one of them is nonzero. (The use of integers is equivalent to that of rationals and more practical.) 

When additional relations such as \re{int_rel} are present, the Kronecker theorem can be formulated as follows \cite{LZ}.

\begin{ver}
Assume that the real parameters $x_s, \: s=0,1,\dots,N$ are all distinct and moreover that there are $L$ relations of the type \re{int_rel} with nontrivial sets $\{r_0^{(i)}, \dots r_N^{(i)} \}$ of integers. 

Then, the approximation condition \re{Kron} holds for every $\delta>0$ if and only if the real quantities $a_i$ satisfy the conditions 
\be
r_0^{(i)} a_0  + r_1^{(i)} a_1 + \dots + r_N^{(i)} a_N \equiv 0  \quad   \mbox{(mod} \;  2 \pi), \quad i=1,2,\dots, L  \lab{ra_cond} \ee
with the same integers $r_i^{(i)}$ as in \re{int_rel}. 
\end{ver}
For the proof of this statement see \cite{LZ}.

Using this (generalized) version of the Kronecker theorem we can formulate the necessary and sufficient conditions for APST.

\vspace{3mm}

{\bf General result.}   {\it Let $x_0, x_1, \dots, x_N$ be $N+1$ distinct eigenvalues of the Jacobi matrix $J$ corresponding to the $XX$ spin chain \re{H_def}. Assume that there are $L \le N$ relations of the type \re{int_rel} with nonzero integer parameters $r_s^{(i)}$. 

Then, the following conditions are necessary and sufficient for APST:

(i) the Jacobi matrix $J$ is mirror-symmetric, i.e. 
\be
B_{s}=B_{N-s}, \; J_s = J_{N+1-s}, \quad s=0,1, \dots N; \lab{BU_mirror} \ee

(ii) the $L$ linear relations 
\be
\sum_{s=0}^N r_s^{(i)}(\pi s - \phi)=0 \quad (\mbox{mod} \; 2 \pi), \quad i=1,2,\dots L  \lab{rel_a_phi} \ee
are compatible.} 

An obvious special case of this statement occurs when all eigenvalues $x_s$ are linearly independent over the field of rational numbers. This means that $L=0$, i.e. that there are no additional relations such as  \re{rel_a_phi} and that version 1 of the Kronecker theorem suffices to conclude to the occurrence of APST . In such instances , we see that the mirror symmetry  of $J$ and the linear independence of the eigenvalues represent the necessary and sufficient conditions for APST.

Let us now compare the conditions for APST with the conditions for PST. It is known that the PST conditions are \cite{Kay}:

(i) the Jacobi matrix $J$ is mirror-symmetric;

(ii) Assuming that the eigenvalues $x_s$ are ordered so that  $x_0<x_1<\dots <x_N$, the differences $\Delta_s = x_{s+1}-x_s$ are proportional to odd numbers:
\be
\Delta_s = \kappa(2j_s+1), \lab{PST_con} \ee  
where $j_s$ are integers and $\kappa$ is a constant not depending on $s$.

We see that mirror-symmetry is necessary for both PST and APST. This implies that the matrix $J$ and hence the Hamiltonian can be uniquely reconstructed from the eigenvalues $x_0, x_1, \dots x_N$ when either process occurs. Indeed, since \re{chi_N_s} (which implies mirror-symmetry) specifies $\chi_N(x)$ at $N+1$ distinct points, this polynomial can be obtained by Lagrange interpolation. Two orthogonal polynomials associated to the 3-diagonal matrix $J$ are therefore known: $\chi_N(x)$ and the characteristic polynomial $\prod_{s=0}^N (x-x_s)$ of degree $N+1$. As shown in \cite{VZ_PST}, all the orthogonal polynomials can then be found by the Euclidean algorithm thereby providing their 3-term recurrence relation explicitly, i.e. giving all the couplings $J_i$ and magnetic fields $B_i$.  

The difference in the conditions for PST and APST arises only in the restrictions on the spectrum or the eigenvalues of J. As is observed, these conditions are much more stringent for PST than for APST.

\section{Spectral surgery and APST}
\setcounter{equation}{0}
A spectral surgery procedure was proposed in \cite{VZ_PST}, as a way to construct $XX$ spin chains with PST from such systems already known to possess the PST property.

Given the initial spectrum $X_N= \{x_0, x_1, \dots, x_N\},$ under spectral surgery, one or several levels  $x_{i_1}, x_{i_2}, \dots, x_{i_j}$ are removed from the set $X_N$.

The reduced set is $\t X_{N-j}=X_N \setminus R_j$, where $R_j = \{x_{i_1}, x_{i_2}, \dots, x_{i_j}\}$ is the set of levels that have been removed. The spectral levels determine the (mirror-symmetric) Jacobi matrix $J$ (and hence the Hamiltonian of the $XX$ spin chain) uniquely. Thus the new spectral set $\t X_{N-j}$ generates a new Jacobi matrix $\t J$ of dimension $(N+1-j) \times (N+1-j)$. Certain restrictions need to be imposed on the possible choices of sets $R_j$. Namely, $R_j$ can contain any number of levels starting from the first one, say $R_{M_1}^{(0)}=\{x_0, x_1, \dots, x_{M_1-1}\}$ or similarly, any number of levels from the last level, say $R_{M_2}^{(N)}=\{x_N, x_{N-1}, \dots, x_{n-M_2+1}\}$. The only restriction is that there should be no gaps in the above sequences. Apart from these two elementary surgeries, levels can also be extracted from the middle of the spectral set $X_N$; in this case the requirement is the following: all such level sets should consist in the union of subsets $R_{L}^{(i)}=\{x_i, x_{i+1}, \dots, x_{i+L-1}\}$ of even length $L$ and without gaps within $R_{L}^{(i)}$. Equivalently, one can say that only pairs $x_i, x_{i+1}$ of neighbor levels can be removed from the middle of the set $X_N$. 

Remarkably, the Jacobi matrix $\t J$ corresponding to a surgered set $\t X_{N-j}$, is obtained from the initial Jacobi matrix by $j$ Christoffel transforms (see \cite{VZ_PST} for details).  

That is, after $j$ Christoffel transforms, we obtain a new Jacobi matrix $\t J$ which is mirror-symmetric and satisfy the PST property. This observation allows to generate new explicit examples with PST without the need to perform the inverse spectral problem algorithm. This is advantageous since the formulas of the Christoffel transform are rather simple.

One such example was already given in \cite{VZ_PST}. Let $N$ be odd. We start with the uniform grid $X_N = \{-N/2, -N/2+1, \dots, N/2\}$ and then remove $2L$ levels symmetrically from the middle of the set $X_N$ 
\ba &&X_{N-2L}=\{-N/2, -N/2+1, \dots
-L-3/2,  -L-1/2,  \nonumber \\ && L+1/2, L+3/2, \dots, N/2-1, N/2 \}.
\lab{gap_x} \ea
Such a spectral set corresponds to the model of the $XX$ spin chain with PST proposed in \cite{Shi}. 

Consider what is the effect of spectral surgery with respect to APST.

It is almost obvious that under an arbitrary admissible surgery procedure the APST property survives. Indeed, if all eigenvalues $x_i$ were linearly independent over the field of rational numbers then any reduced set $\t X_{N-j}$ will obviously satisfy the same property.

Assume that there exists a set of linear relations \re{int_rel}  on the spectral levels $x_s$. APST is possible iff the set of linear relation \re{rel_a_phi} is compatible (it is assumed {\it a priori} that the matrix $J$ is mirror-symmetric).

Under spectral surgery some levels of the set $X_N$ are removed. This means that some of relations \re{int_rel} will disappear (or remain the same). This implies that the new Jacobi matrix $\t J$ will also possess APST. 

Interestingly, this method can sometimes generate a chain with APST, even if the initial chain lacks this property. Indeed, the absence of APST for a given mirror-symmetric $XX$ chain means that relations \re{rel_a_phi} are incompatible. When we remove some of the levels $x_i$ the number of relations  \re{int_rel} can be reduced which in principle can lead to the compatibility of the reduced set of relations \re{rel_a_phi}. Of course one should seek to realize  this  in concrete examples.  We plan to investigate this problem in the future.

\section{Examples and special cases}
\setcounter{equation}{0}
We shall present a number of examples and special cases in this section: 

i) we shall indicate how the central result specializes in the absence of magnetic fields;

ii) we shall record the circumstances for APST in the uniform $XX$ chain and shall discuss the relation between the waiting time and the length of the chain for this particular system, using results  on almost perfect return (to the point of origin);

iii) we shall show, not too surprisingly, that the conditions for PST are a special case of the conditions for APST; 

iv) we shall consider in details the case $N=4$ corresponding to spin chains with 5 sites. In this instance, it is possible to give a very explicit analysis of APST and to estimate waiting times using standard methods of Diophantine approximation. The conditions for all the possible cases (PST, APST and neither) are derived.

v) we shall introduce a remarkable model based on the para-Krawtchouk polynomials that can exhibit  PST, APST or neither depending on the value of one of its parameters ; it will furthermore be seen that the waiting times can be estimated and found to be finite independently of the size of the chain for arbitrarily good fidelities in some special APST situations.

{\bf i. Absence of magnetic fields.}

 Consider the special situation corresponding to zero magnetic fields $B_s=0$. In this case the Jacobi matrix $J$ has only two nonzero diagonals. The corresponding orthogonal polynomials are symmetric $P_n(-x) =-P_n(x)$ and hence the eigenvalues satisfy the properties
\be
x_s + x_{N-s}=0, \quad s=0,1,\dots, N \lab{xx_sym} \ee
where it is assumed that $x_0<x_1 < \dots x_N$.

Consider first the case $N$ odd. Then one has $(N+1)/2$ independent relations \re{xx_sym} for $s=0,1,\dots, (N-1)/2$. They coincide with relations \re{int_rel} where only two integers  are nonzero and given by $r_s=r_{N-s}=1$. 

Relations \re{rel_a_phi} then reduce to the single condition
\be
\pi N -  2 \phi =0 \quad (\mbox{mod} \; 2 \pi)  \lab{cond_phi_odd} \ee
from which we find
\be
\phi =\left\{   {-\pi/2  \quad \mbox{if} \quad  N=4m+3  \atop  \pi/2  \quad \mbox{if} \quad N=4m+1 } \right . \quad   (\mbox{mod} \;  2\pi) . \lab{phi_odd} \ee
If one assumes that the eigenvalues $x_0, x_1, \dots, x_{(N-1)/2}$ are linearly independent, then the Kronecker theorem guarantees with $\phi= \pm \pi/2$ that the corresponding spin chain  will exhibit APST. 

If $N$ is even, we shall have the $N/2$ independent relations \re{xx_sym} for $i=0,1,\dots, N/2-1$ and in addition 
\be
x_{N/2}=0 . \lab{x_0_even} \ee 
Conditions \re{rel_a_phi} are compatible iff
\be
\phi =\left\{   {0  \quad \mbox{if} \quad  N/2 \quad \mbox{is even} \atop  \pi  \quad \mbox{if} \quad N/2 \quad \mbox{is odd}} \right . \quad   (\mbox{mod} \;  2\pi).  \lab{phi_even} \ee
Again, if the eigenvalues $x_0, x_1, \dots x_{N/2-1}$ are linearly independent, the APST property is present.

{\bf ii. The uniform $XX$ chain with zero magnetic fields.}

The APST (or pretty good state transfer) of the $XX$ chain with uniform couplings and no magnetic fields, i.e. $J_s=1$ and $B_S=0$ for all $s$, have recently been sorted out in \cite{Godsil1}. It has been found there that a chain of length $N+1$ admits APST if and only if $N=p-2$ or $2p-2$ where $p$ is prime or iff $N=2^m-2$. As a matter of fact, the proof of that result given in \cite{Godsil1}, effectively proceeds from the application of version 2 of the Kronecker theorem and the use trigonometric relations between the eigenvalues 
\be
x_s= 2 \cos(\pi s/(N+2)), \: s=1,2,\dots,N+1 
\lab{x_s} \ee
which are given by the roots of the Chebyshev polynomials of second kind for this model.

In the same reference \cite{Godsil1} (see also \cite{Godsil}), it is pointed out, on the basis of numerical analysis, that the waiting times for APST will grow with the size of the chain. This has cast doubts on the practical use of APST. Our next examples will hopefully dissipate this negative view by showing that there are models exhibiting APST where the waiting times stay finite as the size is grown.

Before we leave this uniform $XX$ chain, we would like to offer the following comment on the waiting times as the chain is taken to have very large length. 

The quantity
\be
f_{00}(t)= \langle e_0 | e^{-itJ} |  e_0 \rangle \lab{f_00} \ee    
represents the return amplitude, i.e. the amplitude to return to the initial (input) state after a time $t$. Since 
$$
\left |e^{-itJ} |e_0\rangle - e^{i \phi} |e_N \rangle \right | = \left |e^{itJ} |e_N\rangle - e^{-i \phi} |e_0 \rangle \right |,
$$
if we have APST from $| e_0 \rangle$ to $| e_N \rangle$, we also have APST from $| e_N \rangle$ to $| e_0 \rangle$. As a consequence, in a chain with APST, we must observe that there is almost perfect return, that is that there are times $t_n, \: n=0,1,2,\dots$ such that 
\be
\lim_{n \to \infty} |f_{00}(t_n)| =1. \lab{almost_return} \ee
The conditions for almost perfect return have been analyzed in \cite{GVZ} and can be simply stated: An $XX$ spin chain will exhibit almost perfect return if and only if its one-excitation Hamiltonian $J$ has a pure point spectrum. As is readily seen, the spectrum of the uniform $XX$ chain becomes continuous as $N$, the number of sites minus one, becomes infinite (see \cite{GVZ} for more details.) It follows that almost perfect return does not occur in the case of a semi-infinite Heisenberg $XX$ chain and hence not surprisingly an infinite time is required for APST in this model.

{\bf iii. PST as a special case of APST.}

If a set $\{x_0, x_1, \dots x_N\}$ of eigenvalues satisfy the  APST conditions, it is clear that the affine transformed eigenvalues $\t x_s = \alpha x_s + \beta$ will also satisfy the APST conditions. In particular, we can always assume that $x_0=0$ and $x_1=1$. The PST condition \re{PST_con} then implies  that $x_{2s}=K_{2s}$ are even integers while $x_{2s+1}=K_{2s+1}$ are odd integers. We have $N$ linear relations
\be
K_s x_1 =  x_s, \; s=0,2,3,\dots, N .\lab{PST_rel} \ee
With $\phi=0$ we see that the conditions \re{rel_a_phi} become
\be
K_s = s    \mbox{(mod} \;  2). \lab{K_s_PST} \ee 
This is equivalent to the condition that the parity of the integers $K_s$ coincides with the parity of $s$, i.e. with the PST condition. We thus see that the PST condition is a special case of the APST condition, as expected.

A note is in order here. In general the translation $x_s \to x_s + \beta$ of the energy spectrum has no physical meaning: it is always possible to redefine the value of the ground state energy and for instance to put it equal to zero as noted. (Similarly the scale can be arbitrarily chosen using the dilation factor $\alpha$.) This freedom relates to the values that the phase $\phi$ takes modulo $2 \pi$. For a given spectrum, the APST conditions might restrict $\phi$,  however shifts in the ground state energy will allow $\phi$ to take arbitrary values.

{\bf iv. Spin chains with 5 sites and without magnetic field}

Consider the the $XX$ spin chain without magnetic fields ($B_0=B_1 = \dots =B_N=0$) that contains 5 sites. In this case it is possible to express the APST conditions explicitly and to describe 3 possible scenarii: PST, APST and neither.

This corresponds to $N=4$. There are only two distinct exchange constants: $J_1$ and $J_2$. Indeed, due to the mirror-symmetry condition \re{BU_mirror} we have (for $N=4$) that $J_4=J_1, \: J_3=J_2$.

The spectrum of the corresponding Jacobi matrix $J$ can easily be found:
\be
x_0=-a, \: x_1=-b, \: x_2=0, \: x_3=b, \: x_4=a,   \lab{spec_N=4} \ee
where $a=\sqrt{J_1^2+2J_2^2}, \; b= J_1$ are two positive parameters. Clearly $a>b$ so the eigenvalues $x_s$ in \re{spec_N=4} are given in increasing order.

This is a special case of the situation considered in {\bf i.} The sufficient condition for APST is that the parameters $a$ and $b$ be linearly independent over the rationals. This is equivalent to the statement that the ratio $a/b$ is an irrational number. We thus see that if $\sqrt{J_1^2+2J_2^2}/J_1$ is an irrational number, then 
the spin chain with 5 nodes has the APST property.

Suppose now that the ratio $a/b =n/m$ is a rational number where $n,m$ are co-prime integers. This means that there are two conditions on the eigenvalues
\be 
mx_0 -n x_1=mx_4-nx_3=0 \lab{cond_N=4} \ee
Then the relations \re{rel_a_phi} which are required for APST (with $\phi=0$) reduce to the only condition that $n$ be even and $m$ be odd.

But this is equivalent to the PST condition. Indeed, we have  
\be
\frac{a-b}{b}= \frac{n}{m}-1 = M/m,
\lab{PST_N=4} \ee
where $M$ and $m$ are both odd. Observe that the differences between the eigenvalues are $\Delta_1=x_1-x_0=a-b, \: \Delta_2=x_2-x_1=b, \: \Delta_3=x_3-x_2=b, \: \Delta_4=x_4-x_3=a-b$. Hence \re{PST_N=4} coincides with the PST condition for a chain with 5 sites. 
 
In all other cases, i.e. when $n$ is odd and $m$ is odd, or when both $n$ and $m$ are odd, the APST condition is not valid and the $XX$ spin chain does demonstrate neither PST nor APST.

In the APST situation (i.e. when $a/b$ is irrational), it is interesting to estimate the waiting times. It is easy to calculate the amplitude $f_{0N}(t)$ directly from \re{f_0N}:
\be 
f_{0N}(t) = \frac{a^2-b^2}{2a^2} + \frac{b^2}{2 a^2} \: \cos at - \frac{1}{2} \: \cos bt \lab{f_N=4} \ee
This is a simple example of almost periodic functions: two harmonic oscillations with frequencies $\omega_1=a$ and $\omega_2=b$ do not constitute a pure periodic function. But the Kronecker theorem guarantees that there exists a sequence $t_n$ such that $f_{0N}(t_n) \to 1$ when $n \to \infty$.

There are several simple algorithms to estimate the quantities $t_n$.

One of them proceeds through the expansion of $a/b$ into a continued fraction \cite{HW}
\be
z=a/b=\zeta_0 +
{1\over\displaystyle \zeta_1 + {\strut 1 \over\displaystyle
\zeta_2 + \dots }}, \lab{cont_fr_gen} \ee    
where the quotients $\zeta_0, \zeta_1, \dots$ are integers. The corresponding convergents $p_n/q_n$ of the continued fraction are rational numbers which provide the best approximation of the irrational number $a/b$ \cite{HW}. In our case this method works well if all (or at least infinitely many) convergents have the property that $p_n$ is even and $q_n$ is odd. Otherwise one can apply the method of Farey fractions \cite{HW} or, equivalently, the method of intermediate convergents \cite{Lang}.

This method can described as follows. Starting with the convergents $\{p_0/q_0, p_1/q_1, \dots\}$ we compute the so-called mediants (or intermediate convergents) by the formula \cite{Lang}
\be
\frac{\t p_n}{\t q_n} =\frac{p_{n+1} \pm p_n}{q_{n+1} \pm q_n} . \lab{mediant} \ee
The convergents $p_n/q_n$ together with the intermediate convergents  $\t p_n/ \t q_n$ form a set of generalized convergents $w_n=u_n/v_n, n=0,1,2,\dots$, where $u_n=p_n$ or $u_n=\t p_n$ (similarly $v_n=q_n$ or $v_n=\t q_n$).  From the elementary properties of the convergents it follows that all fractions $u_n /v_n$ are simple (i.e. $u_n$ and $v_n$ are co-prime). Moreover, it is possible to show \cite{Lang} that the generalized convergents provide the general solution to the best approximation problem: they satisfy the inequality 
\be
|z-u_n/v_n|<v_n^{-2} \lab{best_uv} \ee 
It is easy to show that there are infinitely many generalized convergents $u_n/v_n$ with the desired property: $u_n$ is even while $v_n$ is odd.

Indeed, assume that there is only a finite number of the convergents $p_n/q_n$ with the desired property. This means that starting with some $n=M$ the numerators $p_n$ are all odd; the corresponding denominators $q_n$ may be either even or odd. Necessarily, for any neighbor pair $p_n/q_n$ and $p_{n+1}/q_{n+1}$, the denominators $q_n$ and $q_{n+1}$ should be of opposite parity. Indeed, if one assumes that $q_n$ and $q_{n+1}$ have the same parity then their mediant  $(p_n + p_{n+1})/(q_n + q_{n+1})$ will be reducible (with both numerator and denominator even) which is impossible. Thus the mediants $\t p_n / \t q_n$ will have the desired property, because $\t p_n =p_{n+1} \pm p_n$ is even as a sum of two even numbers and $\t q_n = q_{n+1} \pm q_{n}$ is odd as a sum of two number with opposite parities. This means that for any irrational number $z=a/b$ there exists an infinite sequence of generalized convergents $c_n=u_n/v_n$ such that $\lim_{n \to \infty}c_n \to z$ and $u_n$ is even while $v_n$ is odd. Moreover, the best approximation property \re{best_uv} holds for all such convergents.

Given this sequence, let us put 
\be
t_n=\frac{\pi v_n}{b}. \lab{t_n_5} \ee 
The amplitude $f_{0N}(t_n)$ then becomes
\be
f_{0N}(t_n)= \frac{a^2-b^2}{2a^2} + \frac{b^2}{2 a^2} \: \cos \left(\pi v_n \frac{a}{b} \right) + \frac{1}{2}  . \lab{f0_tn} \ee
This can easily be simplified to 
\be
f_{0N}(t_n) = 1- \frac{b^2}{a^2} \sin^2 (\pi v_n \ve_n/2), \lab{f0_simpl} \ee
where
\be
\ve_n = \frac{a}{b} - \frac{u_n}{v_n} \lab{epsilon} \ee
is the accuracy of the rational approximation of the irrational number $a/b$. 

By property \re{best_uv} we have that 
\be
f_{0N} \approx 1- \frac{\pi^2 b^2}{a^2 v_n^2} . \lab{f_0_approx} \ee
It is seen from this formula  that $f_{0N}(t_n)$ converges to 1 when $n \to \infty$. Moreover, this formula allows to estimate the accuracy of this approximation if the (generalized) convergents $u_n/v_n$ are known explicitly.  

It is seen that $f_{0N}(t_n) \to 1$ when $t_n \to \infty$. This means that the sequence $\{t_n\}$ is associated to APST.

Consider the special case $J_1=J_2=1$ so that all couplings are now equal. For the homogeneous Heisenberg chain (i.e. with $J_i=1$ for all $i=0,1,2,\dots,N$) it is known \cite{Christ} that PST only occurs when there are 2 or 3 sites. We are thus in a situation of APST for a uniform chain with 5 nodes. 

In this case $a=\sqrt{3}, \: b=1$. The continued fraction representation is
\be \sqrt{3} = 1 +
{1\over\displaystyle 1 + {\strut 1 \over\displaystyle
2 + {\strut 1 \over\displaystyle
1+  \dots }}}. \lab{cont_fr_3} \ee  
The first convergents are
\be
\frac{p_n}{q_n}= \left\{1, 2, \frac{5}{3},  \frac{7}{4}, \frac{19}{11}, \frac{26}{15}, \frac{71}{41}, \dots  \right\}. \lab{conv_3} \ee
It is seen that the first two convergents with the desired property (i.e. an even numerator and an odd denominator) are $2/1$ and  $26/15$. Computing the mediants according to \re{mediant}, we obtain the additional intermediate convergent $12/7$. Hence, the first 3 desired (intermediate) convergents are
\be
\frac{u_n}{v_n} = \left\{ \frac{2}{1}, \frac{12}{7}, \frac{26}{15}    \right\}  \lab{des_inter} \ee
The first fraction, $2/1$, yields an approximation to the difference between $|f_{0N}(t)|$ and 1  with an accuracy of $3 \cdot 10^{-1}$; the second fraction $12/7$ yields an approximation with accuracy $2 \cdot 10^{-2}$.

So, already in the case of a spin chain with 5 sites we see that all 3 possibilities: PST, APST or neither can occur. APST is generic, i.e. APST happens for almost all possible values of $J_1$ and $J_2$.

We shall consider next a $XX$ spin chain with an {\it arbitrary} (even) number of spins which demonstrates the similar property.

{\bf v. Spin chain corresponding to the para-Krawtchouk polynomials.}

Assume that $N$ is odd and that the spectrum $x_s$ is such that (up to an affine transformation) \be x_{2s}=2s, \; x_{2s+1}=2s+\gamma, \; s=0,1,\dots, (N-1)/2 , \lab{exam_no} \ee  
where $\gamma$ is a real parameter such that $0<\gamma<2$.

This situation can be realized in spin chains that are obtained from the para-Krawtchouk polynomials with the recurrence coefficients \cite{VZ_para}
\ba
&&B_n=\frac{\gamma+N-1}{2}, \nonumber \\ &&J_n^2=
\frac{n(N+1-n)((N+1-2n)^2-\gamma^2)}{4(N-2n)(N-2n+2)}.
\lab{para_ub} \ea
The corresponding Jacobi matrix is mirror symmetric. When $\gamma=1$ the grid is uniform and the Jacobi matrix $J$ which is associated to the ordinary Krawtchouk polynomials generates PST \cite{Albanese}. 

It is clear that all eigenvalues can be expressed as linear combinations of $x_1$ and $x_2$. Indeed,
\be 
x_{2s}= s x_{2}, \quad x_{2s+1}=x_1 + s x_2, \quad  s=0,1,2,\dots (N-1)/2 \lab{para_con} \ee
Hence, according to our central result \re{rel_a_phi}, in order to have APST, the following conditions must be realized: 
\be
 a_{2s}= s a_{2}, \quad a_{2s+1}=a_1 + s a_2, \quad   \mbox{(mod} \;  2\pi)   \lab{para_con_a} \ee
where $a_s=\pi s -\phi$. These relations are compatible iff $\phi =0 \quad   \mbox{(mod} \;  2\pi)$. 

If the parameter $\gamma$ is irrational then there are no additional relations for the eigenvalues $x_1, \: x_2$ and thus provided \re{para_con_a} is verified, APST will occur. 

Equivalently, this means that there exists a sequence $\{t_n, \: n=1,2,\dots \}$ such that 
\be
\lim_{n \to \infty}|f_{0N}(t_n)| = 1 \lab{lim_tn} \ee

As in the previous case it is possible to determine  the times $t_n$ and estimate the corresponding rate of convergence in \re{lim_tn} using standard Diophantine approximation methods. . 

Indeed, we can always choose an infinite set of (intermediate) convergents $\{u_1/v_1, u_2/v_2, \dots, u_n/v_n, \dots \}$ with the property that the numerators $u_n$ are all odd.  Introduce the value 
\be
\ve_n =\gamma-\frac{u_n}{v_n} \lab{eps_gamma} \ee
which describes the accuracy of the Diophantine approximation of the parameter $\gamma$. We already saw that 
\be
|\ve_n|<v_n^{-2} . \lab{ve_para} \ee
Choose 
\be
t_n=\pi v_n. \lab{t_n-para} \ee 
Substituting this in  \re{f_0N} and taking into account that $u_n$ is odd we have
\ba
&&f_{0N}(t_n) = -\sum_{s=0}^{(N-1)/2} w_{2s} + \sum_{s=0}^{(N-1)/2} e^{-i \pi \gamma v_n } w_{2s+1} = \nonumber \\
&&-\sum_{s=0}^{(N-1)/2} w_{2s} - e^{-i \pi v_n \ve_n} \sum_{s=0}^{(N-1)/2}  w_{2s+1} . \lab{A_t_n} \ea
In \cite{VZ_para} it was shown that for the para-Krawtchouk polynomials the identities
\be
\sum_{s=0}^{(N-1)/2} w_{2s} = \sum_{s=0}^{(N-1)/2} w_{2s+1} =1/2 \lab{id_para} \ee
hold. Hence
\be
f_{0N}(t_n)= -\frac{1}{2} \left( 1+ e^{-i \pi v_n \ve_n}\right).\lab{f_0N_para} \ee
Now
\be
|f_{0N}(t_n)| = \cos (\pi v_n \ve_n/2) \approx 1 - \frac{\pi^2}{8v_n^2} \lab{f_para} \ee
where we have used \re{ve_para}.

Formula \re{f_para} gives a good approximation of the amplitude $|f_{0N}|$ if $v_n$ is sufficiently large (depending on the physical requirements on the accuracy). E.g. for $v_n>10$ we get an accuracy of $1\%$.

Consider, e.g. the value $\gamma=\sqrt{3}$. 

From \re{cont_fr_3} we find the first appropriate (i.e. with all numerators $u_n$ odd) convergents 
\be
\frac{u_n}{v_n} = \left\{1, \frac{5}{3}, \frac{7}{4},  \frac{19}{11}, \frac{45}{26}, \frac{71}{41}, \dots, \right\} . \lab{con_3_para} \ee
 Already the third convergent $7/4$ gives an accuracy of about $1\%$. This is because the actual accuracy $\ve_3  \approx 0.0179$ of the third convergent is smaller than what the rhs of formula  \re{ve_para} gives. Hence the first waiting time corresponding to an accuracy of $1\%$ in the amplitude is $t_3=4\pi$.

Consider now the case where $\gamma=p/q$ is a rational number, with $p$ and $q$ coprime integers. In this case there is an additional linear relation between $x_1$ and $x_2$:
\be
2q x_1 = p x_2 \lab{add_para_x} \ee  
and hence a similar relation should hold for $a_1$ and $a_2$:
\be
2q a_1 = p a_2 \quad   \mbox{(mod} \;  2\pi)  \lab{add_para_a} \ee 
where $a_1 =\pi, \: a_2 = 2 \pi .$
If the numerator $p$ is odd then  \re{add_para_a} holds for any pair of coprime integers $p,q$ and hence the APST condition is satisfied. In fact we are then in the situation where not only APST but even PST occurs. Indeed, this corresponds to the model with PST that we derived and discussed in \cite{VZ_para}.

If the numerator $p=2j$ is even for some integer $j$ (in this case, necessarily the denominator $q$ is odd) then relation \re{add_para_x} becomes $q x_1 = j x_2$ and hence  \re{add_para_a} reads
\be
q \pi - 2 j \pi =0  \quad \mbox{(mod} \;  2\pi).  \lab{add_para_even} \ee 
Obviously, \re{add_para_even} cannot hold when $q$ is odd and there is no PST nor APST in this case.

In summary the picture is as follows:

(i) when $\gamma$ is an irrational number, APST is observed;

(ii) when $\gamma=p/q$ is rational with $p$ odd there is PST;

(iii) when $\gamma=p/q$ is rational with $p$ even neither PST nor APST happens.  

Let us make the following remark in connection with the last two examples. As mentioned in subsection iii, in the discussion of the uniform $XX$ chain, almost perfect return occurs in chains of the type we have been considering, this is so whenever the spectrum  of the Jacobi matrix $J$ is discrete. Clearly, this will be the case for any $XX$ spin chain with nearest-neighbor interactions that has a finite number of sites. It should be stressed that finite length does not similarly imply APST; as we have seen, there are examples of systems with finite 1-excitation Hamiltonian $J$ that do not possess  APST.

\section{Conclusions}
\setcounter{equation}{0}
Let us recapitulate the essential elements of our analysis of almost perfect state transfer in spin chains. It builds on the knowledge that $XX$ spin chains with properly engineered couplings and magnetic fields can effect the transport of states from one end to the other with probability 1 over certain times - these are chains with perfect state transfer (PST). In view of the fact that there are always manufacturing or measurement imprecisions, our study aimed to characterize the models where although not perfect, state transfer could be realized with a probability very close to 1 - such chains have been said to show almost perfect state transfer (APST).

We have thus undertaken to categorize all $XX$ spin chains with arbitrary nearest-neighbor couplings and magnetic fields that would exhibit APST. The Kronecker theorem in Diophantine approximation proved essential in this investigation. The necessary and sufficient conditions that have been found for a chain to admit APST bear on the 1-excitation restriction $J$ (a tri-diagonal matrix) of the chain Hamiltonian. Not unlike what is needed for PST, the requirements for APST entail a symmetry condition on $J$ as well as spectral restrictions. As it turns out, like for PST, $J$ must be mirror-symmetric that is, invariant against reflection with respect to its anti-diagonal. The conditions on the eigenvalues of $J$ turn out  (in general)  to be much less demanding for APST than for PST. They amount roughly to the conditions for which the Kronecker theorem applies and imply that at least a subset of these eigenvalues be linearly independent over the field of the rational numbers. Spin chains with APST therefore much enlarge, in principle, the class of such systems that can be exploited as quantum wires since their fidelity can be as good as is technically relevant.

It has also been demonstrated how chains with APST can be constructed from a chain that already admits (or possibly not) APST, by the removal of single excitation energy levels from the parent system via Christoffel transforms. This offers a constructive method to broaden the catalog of systems with APST.

A number of examples and special cases interesting in their own right have also been presented and discussed. They illustrate situations where PST or APST occur or where neither can happen. One of the models introduced have provided counter-examples to the view borne out in particular from the examination of the Heisenberg  $XX$ chain with homogeneous couplings, that APST requires excessively long times in extended wires.

We trust this report provides interesting information on the transfer of state in $XX$ spin chains with nearest-neighbor couplings and suggests that APST requires further analysis.

\bigskip\bigskip
{\Large\bf Acknowledgments}
\bigskip

AZ thanks Centre de Recherches Math\'ematiques (Universit\'e de
Montr\'eal) for hospitality.  The authors would like to thank
M.Christandl, M.Derevyagin and A.Filippov for stimulating
discussions. They also acknowledge communications from M.Bruderer, D. Burgarth and S. Severini and are also very grateful to a referee for a very constructive and stimulating review.
The research of LV is supported in part by a research grant from the Natural Sciences and Engineering Research Council
(NSERC) of Canada.

\newpage

\bb{99}

\bi{Albanese} C.Albanese, M.Christandl, N.Datta, A.Ekert, {\it
Mirror inversion of quantum states in linear registers},  Phys.
Rev. Lett. {\bf 93} (2004), 230502.

\bi{besic} A.S.Besicovitch  {\it Almost periodic functions}, (Dover, 1954)

\bi{Bose} S.~Bose, {\it Quantum communication through spin chain
dynamics: an introductory overview},  Contemp. Phys., {\bf 48},
(2007), 13 -- 30.

\bi{Bruderer} M. Bruderer, K. Franke, S. Ragg, W. Belzig, D. Obreschkow, {\it Exploiting boundary states of imperfect spin chains for  high-fidelity state transfer}, Phys. Rev. A {\bf 85} (2012), 022312.  arXiv:1112.4503

\bi{Burgarth} D.Burgarth, {\it Quantum State Transfer with Spin Chains}, arXiv:0704.1309.

\bi{BB} D.Burgarth S.Bose, {\it Conclusive and arbitrarily perfect quantum state transfer using parallel spin chain channels}, Phys. Rev. A
{\bf 71}, (2005) 052315 . arXiv: quantum-ph/0406112v4.

\bi{bounds} C.Burrell and T. Osborne, {\it Bounds on information propagation in disordered quantum spin chains}, Phys. Rev. Lett. {\bf 99}, (2007), 167201  ArXiv: quant-ph/0703209v3

\bi{disorder} C.Burrell, J.Eisert and T.Osborne, {\it Information propagation through quantum chains with fluctuating disorder}, Phys. Rev. A {\bf 80} (2009), 052319 . ArXiv: 0809.4833v1.

\bi{fractal} G. De Chiara, D. Rossini, S.Montenegro and R. Fazio, {\it From perfect to fractal transmission in spin chains}, Phys. Rev. A {\bf 72}, 012323 (2005) arXiv: quant-ph/0502148v2.

\bi{Chi} T. Chihara, {\it An Introduction to Orthogonal
Polynomials}, Gordon and Breach, NY, 1978.

\bi{Christ} M. Christandl, N.Datta,Tony C. Dorlas, A.Ekert, A.Kay and A. J. Landahl, {\it Perfect transfer of arbitrary states in quantum spin networks}, Phys. Rev. A {\bf 71} (2005), 032312. arXiv:quant-ph/0411020

\bi{Godsil} C.Godsil, {\it State Transfer on Graphs}, Discrete Math. {\bf 312}(1): 129--147 (2012). arXiv:1102.4898

\bi{Godsil1} C.Godsil, S.Kirkland, S.Severini, J.Smith, {\it Number-theoretic nature of communication in quantum spin chains}, Phys. Rev. Lett. {\bf 109} (2012), 050502; arXiv:1201.4822.

\bi{GVZ} A.Gr\"unbaum, L.Vinet and A.Zhedanov, {\it Birth and
death processes and quantum spin chains}, arXiv:1205.4689.

\bi{HW} G.H.Hardy and E.M.Wright  {\it An introduction to the theory of numbers}, 6th ed. Oxford University Press, 2008

\bi{Kay1} A.Kay, {\it Perfect State Transfer: Beyond Nearest-Neighbor Couplings}, Phys. Rev. A {\bf 73} (2006), 032306 . ArXiv: quant-ph/0509065v2.

\bi{Kay} A.Kay, {\it A Review of Perfect State Transfer and its
Application as a Constructive Tool}, Int. J. Quantum Inf. {\bf 8}
(2010), 641--676;  arXiv:0903.4274.



\bi{Lang}. S.Lang, {\it Introduction to Diophantine Approximations}, Springer-Verlag, 1991.

\bi{LZ} B.M.Levitan and V.V.Zhikov, {\it Almost periodic functions and differential equations}, Cambridge University Press, 1982.

\bi{MKE} C.Marletto, A.Kay and A.Ekert, {\it How to counteract systematic errors in quantum state transfer}, arXiv: 1202.2978v1.

\bi{Shi} T. Shi, Y.Li , A.Song, C.P.Sun,  {\it Quantum-state
transfer via the ferromagnetic chain in a spatially modulated
field}, Phys. Rev. {\bf A 71} (2005), 032309, 5 pages,
quant-ph/0408152.

\bi{RSA} R.Ronke, T.Spiller, I.D'Amico, {Long-range interactions and information transfer in spin chains}, J. Phys.: Conf. Ser. {\bf 286} (2011), 012020.  arXiv: 1101.4509v1



\bi{VZ_PST} L.Vinet, A.Zhedanov, {\it How to construct spin chains with perfect spin transfer}, Phys.Rev. {\bf A 85} (2012), 012323.

\bi{VZ_para} L.Vinet and A.Zhedanov, {\it Para-Krawtchouk polynomials on a bi-lattice and a
quantum spin chain with perfect state transfer}, arXiv:1110.6475v2.

\eb

\end{document}